# Sensor for monitoring plasma parameters

Alexander A. Bol'shakov,[*] Brett A. Cruden,[†] Surendra P. Sharma

NASA Ames Research Center, Moffett Field, CA 94035, USA

**ABSTRACT**

A spectrally tunable VCSEL (vertical cavity surface-emitting laser) was used as part of sensing hardware for measurements of the radial-integrated gas temperature inside an inductively coupled plasma reactor. The data were obtained by profiling the Doppler-broadened absorption of metastable Ar atoms at 763.51 nm in argon and argon/nitrogen plasmas (3, 45, and 90% $N_2$ in Ar) at pressure 0.5-70 Pa and inductive power of 100 and 300 W. The results were compared to rotational temperature derived from the $N_2$ emission at the (0,0) transition of the $C^3\Pi_u$–$B^3\Pi_g$ system. The differences in integrated rotational and Doppler temperatures were attributed to non-uniform spatial distributions of both temperature and thermometric species ($Ar^*$ and $N_2^*$) that varied depending on conditions. A two-dimensional, two-temperature fluid plasma simulation was employed to explain these differences. This work should facilitate further development of a miniature sensor for non-intrusive acquisition of data (temperature and densities of multiple plasma species) during micro- and nano-fabrication plasma processing, thus enabling the diagnostic-assisted continuous optimization and advanced control over the processes. Such sensors would also enable tracking the origins and pathways of damaging contaminants, thereby providing real-time feedback for adjustment of processes. Our work serves as an example of how two line-of-sight integrated temperatures derived from different thermometric species make it possible to characterize the radial non-uniformity of the plasma.

## 1. INTRODUCTION

A wide variety of plasma-based processes have been developed in the microelectronics industry. These techniques (plasma etching, plasma-enhanced chemical vapor deposition, ion implantation, sputtering, ashing, dry cleaning, *etc*.) are being applied for synthesis and treatment of micro- and nano-scale structures. New plasma technologies are also being developed, in particular, for fabrication and functionalization of carbon nanotube-based devices. The latest trends in development of nanoelectronics have clearly indicated that further progress in this field is contingent upon creating means to incorporate the nano-structured architectures in larger CMOS-platform micro-configurations.[1-5] Thus, development of the plasma treatment techniques applicable to both nano and micro realms (in both top-down and bottom-up fabrication technologies) is necessary.

Single impurity atoms or molecules change drastically the local electronic properties of the nanostructures. Such changes may be purposely used in nano-scale design but may also be extremely damaging in applications where defect-free structures are desirable as the size of a "killer" defect falls below one nanometer (<10% the size of the device). Anchoring specific molecular groups to the structural surfaces is the major way to design and functionalize the nanotube-based systems.[6,7] Metals for nanotube doping (Cs, K, Li, Rb) and/or residues of common catalysts (Co, Ni, Fe used in the nanotube synthesis) can damage transistor gates if inadvertently transferred into the gate area during other fabrication and assembly steps. Hence, atomistic control over dopant placement and contamination monitoring is required.

The probability of contamination and product susceptibility to damaging contamination are multiplying with the increase in functionality and density of integration. The most problematic contaminants for nano- and microelectronic fabrication are atomic metals, residual moisture and oxygen, volatile organic molecules, and nano-size particulates. Continuous *in situ* monitoring of reaction kinetics, local temperature values and physical plasma parameters; determination of a multitude of species; and understanding of the paths of low-level contamination that occurs in the fabrication process are necessary to enable fine optimization through a sensor-driven feedback with high sensitivity and sharp spatial resolution.

---

[*] On leave from V.A. Fock Institute of Physics, St. Petersburg State University, St. Petersburg, 198904, Russia; <alexandb@mail.ru>
[†] Also at Eloret Corporation, Moffett Field, CA 94035, USA; <bcruden@mail.arc.nasa.gov>

Diode laser based optical diagnostics are ideal for non-intrusive *in situ* measurement of temperature in the vicinity of the nano/micro-structures and determination of densities of important plasma species responsible for fabrication or functionalization of devices and detection of damaging contaminants. This can be realized by profiling Doppler widths of absorption lines and by using tunable laser based absorption spectrometry with signal enhancement in an actively modulated high-finesse optical cavity.

Analytical hardware and methodology required for these techniques are simple, straightforward and robust. The technology possesses a unique combination of qualities necessary for the *in situ* ultrasensitive determination of the micro-scale local parameters in real time. It is also capable of continuous measurements of the physical plasma parameters such as electron density and temperature, gas temperature and velocity, and absolute densities of various reacting species. The multiplexing capacity of the diode lasers can be utilized for the comprehensive multi-parametric data acquisition by means of a single, simple, compact, and inexpensive tool.

## 2. ABSORPTION SPECTROSCOPY WITH DIODE LASERS

In nano- and micro-fabrication, temperature measurement around the developing features is of great interest. Synthesis quality is affected by the local microenvironment where temperature plays a distinct role. Inherent to diode lasers, excellent spectral resolution allows examination of the lineshapes that yield both Gaussian and Lorentzian components.[8] As a result, the gas kinetic temperature is derived from the Doppler width; electron number density can be derived from the Stark broadening. The Stark centerline shift is indicative of the electron temperature. A laser beam can be easily adjusted to analyze the immediate vicinity of the growing nanostructures (or features etched down) in real time.

### 2.1 Continuous-wave cavity ring-down spectroscopy

Application of the diode laser based absorption, enhanced in a modulated high-reflectivity optical resonator is a technique in principle capable of recording absorption from single atoms or molecules at strong spectral lines and fundamental bands.[9] It can also characterize nano-size particulates in the plasma reactor. This technique is based on the measurement of the rate of absorption of laser radiation confined in a high-quality resonator and may be referred as continuous-wave cavity ring-down spectroscopy (cw-CRDS). The ultimate sensitivity of CRDS results from both its immunity to fluctuations of the laser intensity and a very long absorption pathlength (up to ~10 km) within the cavity. Calibration at the absolute scale follows directly from the Beer-Lambert law. The diode laser induced fluorescence[10-13] can be simultaneously recorded, which adds three-dimensional resolution.

The cw-CRDS is significantly more sensitive at much higher spectral resolution, relative to other non-intrusive diagnostics (emission, UV absorption, FTIR). Non-optical diagnostics require sampling or extraction from plasma and, hence, are intrusive. CRDS can resolve many of the isotopic molecules and is unaffected by isobaric mass interferences that are so ubiquitous and problematic in mass-spectrometry.

The wavelength range covered by commercially available diode lasers that require no cryogenic cooling extends currently from 0.3 to 2.8 μm and will be widening in future. Within this range, many atomic and virtually every inorganic molecular species can be probed at their electronic or vibrational transitions. Hence, CRDS with diode laser sources is an almost universal technique for monitoring plasma species, temperature and other parameters. Determination of a variety of elements by atomic absorption with the tunable diode lasers have been reported (*e.g.*, Al,[14] Ba,[15] Ca,[16] Cr,[17] Cs,[18] Cu,[19] Gd,[20] Hg,[21] I,[22] In,[23] K,[24] La,[25] Li,[26] Mn,[27] Pb,[24] Rb,[11,18] Sm,[28] U,[29] Y,[30] Zr[31]) and many more listed[32] as possible at doubled laser frequency: Ag, Co, Eu, Fe, Ga, Hf, Ho, Lu, Mo, Nb, Nd, Ni, Os, Re, Rh, Ru, Sc, Tb, Th, Ti, Tl, Tm, V, and W. Densities of common molecular contaminants and major plasma constituents, such as $H_2O$, OH, $O_2$, CO, $CO_2$, $CH_4$, $NH_3$, HCl, and HBr have also been measured by diode laser based absorption.[33-39] Measurement of the $CF_2$ and $CF_4$ etching-related molecules at combination bands $(3\nu_1+\nu_3)$ and $(\nu_1+3\nu_3)$ around 2.12 μm by cw-CRDS has been suggested.[40] An array of species $C_nH_m$ (n,m=0,1,...4) important for nanotube growth[41] can be resolved at their vibrational overtones ($2\nu_3$=1.7μm, $3\nu_3$=1.14μm, $4\nu_3$=0.87μm).

We consider here only those diode lasers that operate at room temperature. Substantial research in plasma etching diagnostics has been carried out with lead-salt diode lasers that operate in mid-infrared but have to be cryogenically cooled. Measurements of $CF_X$, $C_2F_6$, $CHF_3$, FCN, $CH_3$, $CH_4$, $C_2H_X$, $CH_3OH$, $SiH_X$, $SiF_4$, $SF_6$, $BCl_3$ and atomic Cl have been performed in the 7-17μm region for research purposes.[42-49] However, lead-salt laser spectrometers are cumbersome, expensive, and impractical for *in situ* process diagnostics on commercial tools. Most recently, quantum cascade lasers have been developed to produce single-mode mid-infrared radiation at near-room temperature maintainable by Péltier coolers. The latter lasers are only now approaching markets.

Our goal is to meet most of the diagnostic needs in nano- and micro-scale fabrication technologies with a single, robust, cost-effective sensor that will enable the diagnostics-assisted process with continuous refined optimization, temperature and contamination control, and improved reproducibility and reliability. The basic hardware includes a diode laser, a photodiode, two mirrors and a mirror piezo-mount. Modulation and registration electronics do not have to be adjacent to the plasma chamber.

**2.2 Measurements of excited and metastable atom densities and temperature**

Details of driving mechanisms in plasma reactors for fabrication of microelectronics are not fully known even in the case of argon plasma. At pressures between 10 Pa and 10 kPa, metastable atoms play an important role in stepwise excitation and ionization processes[50,51] but direct ionization often dominates in argon plasmas at lower and higher pressures.[52,53] Mechanisms vary depending on regime of plasma operation. Therefore, continuing efforts on plasma diagnostics are warranted.

Diode lasers have been actively used for measuring densities of metastable and emitting excited states of neutral or ionic species of argon, krypton and xenon.[54-60] In addition, they have been successfully used for determination of the gas kinetic (translational) temperature in the rare gas plasmas.[52,55-61] Concentrations of excited atoms of chlorine, fluorine and oxygen have also been measured with diode lasers.[62,63]

Most of previous work with the diode lasers was performed using telecommunication lasers that emit a single spectral mode (no broader than $10^{-4}$ nm) formed in the Fabry-Pérot cavity defined by the laser edge facets. Generally, the Fabry-Pérot diode lasers are capricious in regard to frequent and often uncontrollable mode jumping. A single-mode tunability range of such lasers is narrow (0.03-0.3nm) but significantly wider than Doppler width of spectral lines. Relatively recently, the distributed feedback (DFB) lasers and vertical cavity surface-emitting lasers (VCSEL) have been developed. The devices of both latter types can operate in a single longitudinal mode and are continuously tunable within one or several nanometers without mode jumping. This quality makes them very convenient for spectroscopic applications. Recent investigations[64-68] have shown that VCSEL devices are especially advantageous due to their wide tunability, good linearity and fast response.

## 3. EXPERIMENTAL

The immediate objective of this work was to demonstrate the extent of ease in deriving the important plasma parameters by means of a miniature diode laser. We used a VCSEL diode (Specdilas V-770, Laser Components GmbH) for absorption profiling at the 763.51-nm argon line. The experiments resulted in line-of-sight integrated values of the temperature of argon metastable atoms in the center of the discharge. Our plasma reactor[69] was a modified (inductive) Gaseous Electronics Conference (GEC) reference cell with a grounded bottom electrode. Plasmas in argon and argon/nitrogen mixtures were investigated. The feed gas flow rate was fixed at 8 cm$^3$/min. Pressure was varied from 0.5 to 67 Pa. Incident RF power applied was 100 or 300 W with reflected portion below 1%.

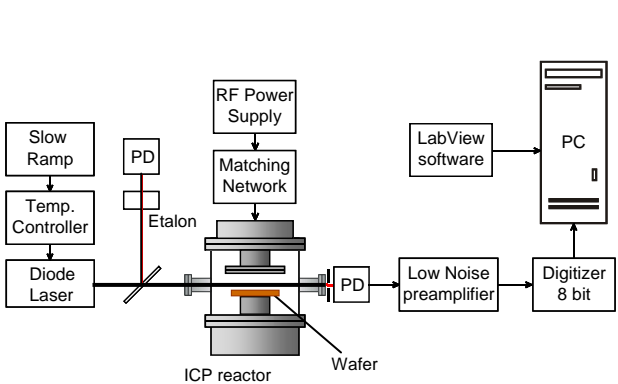 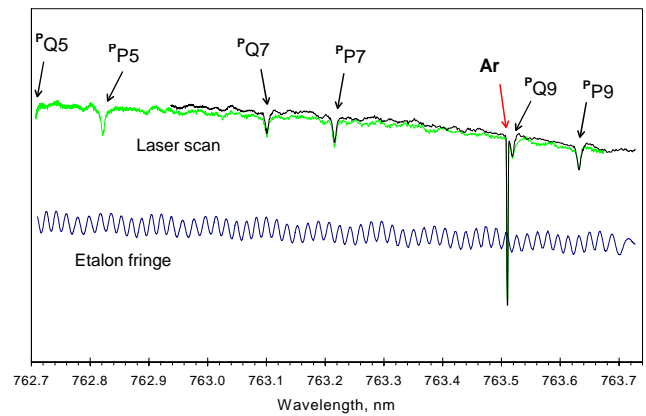

Fig. 1. Simplified diagram of the experimental setup.   Fig. 2. Absorption by argon plasma and ambient air (oxygen).

No absorption enhancement in a reflective cavity was necessary for this purpose. Neither mirrors nor piezo-mounts were installed. A laser beam passed through standard quartz windows of the GEC cell (Fig. 1). A 0.3-meter SpectraPro 300i spectrometer (Acton Research) and SpectruMM back-illuminated CCD camera (Roper Scientific) were used instead of a photodiode, so that we were able to obtain absorption and emission data from plasma simultaneously. The entrance slit of the spectrometer was 50 μm for emission registration but was wide open (0.5 mm) during absorption measurements.

The SpectraPro spectrometer and CCD camera were used to resolve emission spectra for our current research purpose but they would be unnecessary for industrial process diagnostics. A photodiode with an optical filter is sufficient for absorption measurements as spectral resolution in this case is determined by the spectral width of laser radiation, which is very narrow ($1 \times 10^{-5}$ nm with the laser we used).

The diode laser was repeatedly scanned across the argon line at 763.51 nm by changing its case temperature from 29° to 23° C at constant anode current 4.9 mA. Optical power of the laser was about 0.4 mW. Power density was in the range of 0.3–5 mW/cm$^2$ and varied due to beam divergence as it propagated through the plasma volume. Only the central portion of the beam was recorded. Dimensions of the recorded part were limited vertically by 2 to 10 pixels of the detector (0.05–0.24 mm) and 0.5 mm horizontally by the spectrometer entrance slit. No axial (vertical) or lateral spatial profiles across the plasma reactor were attempted at this time. To facilitate wavelength calibration, low-finesse étalon fringes from an ordinary 11-mm thick quartz "window" were simultaneously recorded by a silicon photodiode (see Fig. 2).

For each laser scan, absorption data is first calibrated versus time by use of the étalon trace. Etalon maxima and minima are extracted by performing a quadratic fit in the vicinity of each extremum. The fringe points versus time are then fit to a fourth order polynomial approximation to extract laser wavelength versus time. This fitting procedure was independently verified by comparison to ambient-air oxygen absorption lines (Fig. 2) and readings of the spectrometer calibrated against the argon/mercury lamp spectrum. The uncertainty in relative wavelength calibration from this method is estimated at < 4% and is less than the overall error in the data obtained. The absorption profile of the argon line is fit with a quadratic baseline correction for modulation of the absolute laser intensity in the vicinity of the absorption. At low pressure (<100 Pa) and low electron density (~$10^{11}$cm$^{-3}$), collisional broadening is expected to be small. Fits attempted with a Voigt lineshape indicated a negligible Lorentzian component and produced nearly identical fits to the Doppler profiles. The gas kinetic temperature is then extracted from the half-width at half-maximum (HWHM) of the line broadening obtained in the Doppler fit:

$$T_{Doppler} = \left( \frac{HWHM}{\lambda_0} \right)^2 \frac{c^2 m_{Ar}}{2k \ln 2} \; , \tag{1}$$

where $\lambda_0$ is the wavelength of the argon line (763.51 nm), $c$ is the speed of light, $m_{Ar}$ is mass of argon atom, and $k$ is Boltzmann constant. The reported temperature values represent the average of 4–8 consecutive scans. The standard deviations due to random scatter varied by 15 to 60 K depending on experimental conditions. However, based on the standard deviation of the fit parameters, the temperature would typically be accurate to within a few Kelvin degrees and thus, precision could be improved.

The rotational temperature is extracted from fitting the (0,0) vibrational transition of the 2$^{nd}$ positive system (C$^3\Pi_u$–B$^3\Pi_g$) of N$_2$. The fit procedure described previously,[70] is applied here using the spectral constants of Roux *et al.*[71] and Hamiltonian of Brown and Merer,[72] with the lambda doubling constants removed (as this effect is well below the resolution of the spectrometer). The differences between synthetic spectra and experimental spectra are then minimized with respect to intensity and rotational and vibrational temperatures. Several overlapping argon lines are also accounted for in the fit. The errors for rotational temperature reflect the standard deviation in the parameters from the minimization procedure. When multiple spectra were collected and fit under the same conditions, the results were found to agree within this error.

## 4. RESULTS AND DISCUSSION

The argon line at 763.51 nm is very close to the molecular oxygen line $^P$Q9 of the (0,0) b$^1\Sigma_g$–X$^3\Sigma_g$ transition at 763.52 nm. Hence, we conveniently used the rotational structure of ambient atmospheric oxygen for verification of wavelength calibration. Oxygen absorption was barely seen in most of our experiments when the spectrometer entrance was facing directly the plasma reactor window. The two absorption traces presented in Fig. 2 were recorded at increased pathlength (total 4 m) of the laser beam passing through air between plasma reactor and detector. One of the traces was scanned slower, the other was faster. The latter scan captured five oxygen lines starting from the longer wavelengths ($^P$P9, $^P$Q9,

$^PP7$, $^PQ7$, $^PP5$) and touched a wing of the $^PQ5$ line at the shortest wavelength of the scan. Wavelengths and wavenumbers of these lines along with the argon line are listed in Table 1 (spectroscopic data for oxygen is from Cheah *et al.*[73]).

Additionally, Fig. 2 depicts the low-finesse étalon fringe that was utilized for routine wavelength calibration. We used one of the higher transverse modes formed in the optical leg between the étalon, a quartz beam splitter, and a photodiode in order to avoid excessive modulation observed in the $TEM_{00}$ mode as a result of interfering reflections from the front window of the GEC cell. A fringe period (free spectral range of the étalon) at the higher transverse mode is smaller than that in the $TEM_{00}$ mode. However, some long-period modulation because of the reflections at the GEC cell window can still be noticed in this higher mode fringe (see Fig. 2). The intensity of the laser output was increasing slightly as it was cooled down by Péltier thermoelectric modules during the scans.

Table 1. Oxygen and argon lines scanned.

| Line | Wavelength, nm | Wavenumber, cm$^{-1}$ |
|---|---|---|
| $O_2$ $^PP9$ | 763.63 | 13091.71 |
| $O_2$ $^PQ9$ | 763.52 | 13093.65 |
| Ar | 763.51 | 13093.79 |
| $O_2$ $^PP7$ | 763.22 | 13098.85 |
| $O_2$ $^PQ7$ | 763.10 | 13100.82 |
| $O_2$ $^PP5$ | 762.82 | 13105.62 |
| $O_2$ $^PQ5$ | 762.70 | 13107.63 |

Emission intensity at several argon lines was recorded as a function of pressure in the pure argon discharge at 100 W of incident power. These intensities normalized at the magnitude of their maxima are illustrated in Fig. 3. At lower pressures, emission increases with pressure because both argon and electron densities increase. Then after reaching a maximum, emission decreases because the electron temperature (and the excitation rate) decreases at higher pressures. The higher the energy of the upper level of the emitting line, the more sensitive this line is to the decrease in electron temperature, *i.e.* its emission maximum shifted to lower pressure. An estimate of the electron temperature can be inferred from measurements of relative intensities of highly excited argon lines as their excitation temperature[74]. At our experimental conditions, the electron temperature was between 20000 and 52000 K (2–4.5 eV) in argon plasma[69]. Electrons cannot effectively heat the heavy particles (atoms and ions) and hence, the gas kinetic temperature remains about two orders of magnitude lower. No information on the gas temperature can be obtained from low-resolution atomic emission data, such as presented in Fig. 3. However, it is the gas temperature that is one of the most important parameters in material processing on industrial plasma tools.

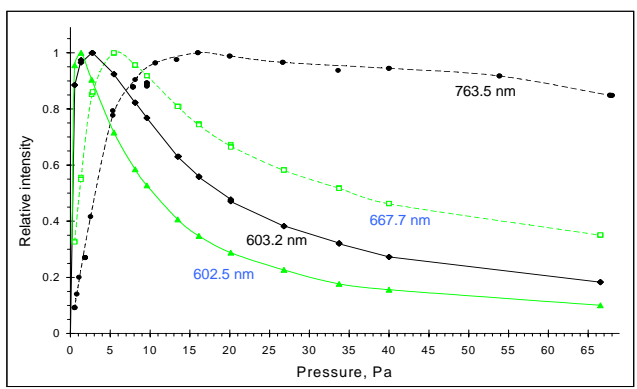 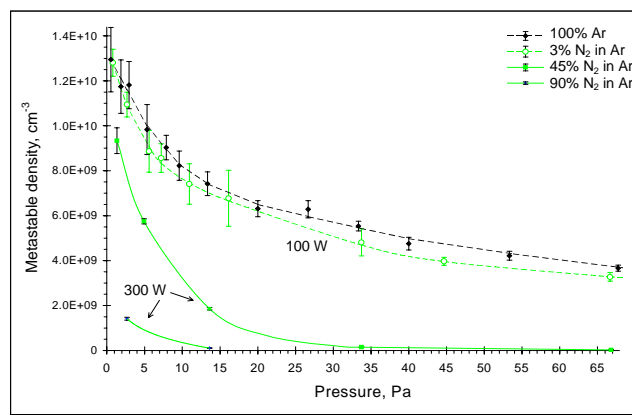

Fig. 3. Emission line intensities from argon plasma at 100 W.      Fig. 4. Number densities of metastable argon atoms.

Line-of-sight integrated number densities of metastable argon atoms as measured by diode laser absorption at 763.51 nm are shown in Fig. 4 depending on pressure in 100%-argon and nitrogen/argon mixtures. Absorption pathlength was assumed equal to the plasma reactor internal diameter of 25 cm. The glowing part of the plasma with high electron density is limited roughly by the diameter of the induction coil (10 cm) but the metastable atoms diffuse from the center and decay on the walls. A small portion of the metastables might protrude even farther into the window ports of the GEC cell (additional distance of 12.5 cm toward the windows on each side of the plasma reactor). The latter portion was neglected in calculation of metastable number densities. The effects of stimulated radiation and fluorescence were also neglected.

At lower pressures, there is no significant difference within random scatter between metastable argon densities in pure argon and 3% nitrogen/argon mixture. At higher pressures, metastable argon densities are lower when nitrogen is added. Apparently, this is because of reduction in electron density resulting from nitrogen introduction and the higher rate of argon quenching by nitrogen molecules than by argon itself. The dependence on pressure of the metastable atoms should, in principle, be similar to that of the emitting atoms (see Fig. 3). A maximum of the metastable density on the pressure scale was not observed here because we did not probe pressures below 0.5 Pa. A comparison of data in Figures 3 and 4 indicates that the bulk of argon plasma emission is largely determined by stepwise excitation (from the metastable states) and cascade transitions from higher states rather than by direct excitation from the ground state. Only the very highly excited emitting Ar atoms behave somewhat similar to the metastables, *e.g.* the line at 602.5 nm that originates from the energy level at 123882.20 cm$^{-1}$ (30738.44 cm$^{-1}$ above the lowest metastable level at 93143.76 cm$^{-1}$).

Nitrogen added in argon plays at lower pressures a simple dilution role. The number density of Ar metastables in the 45% $N_2$/55% Ar mixture at 2.7 Pa was approximated as $7.8 \times 10^9$ cm$^{-3}$, which is exactly 5.5 times higher than $1.4 \times 10^9$ cm$^{-3}$ measured in the 90% $N_2$/10% Ar mixture at the same pressure and power (see Fig. 4). However, the metastable argon density in the 90% $N_2$/10% Ar mixture was more that 3 times lower than the density which would be determined by a dilution factor when compared to the 45% $N_2$/55% Ar mixture at 13.6 Pa. This is again because of quenching of argon metastables by nitrogen molecules that becomes more frequent as pressure increases.

The results of the gas temperature determination from Doppler broadened profiles of argon absorption and from rotational spectra of the molecular nitrogen emission are shown in Figures 5 and 6. All species in the plasma except electrons are assumed to be in translational equilibrium at the gas kinetic temperature. Experimentally derived rotational and Doppler data resulted in effective temperature values, integrated over the line of sight and weighted by the spatial density distributions. These distributions were different for different species. Obviously, the emitting nitrogen molecules existed only in the glowing part of the plasma, while the metastable argon atoms had a broader distribution filling out the dark space near the reactor walls. Thus, the Doppler broadening measurements yield an average gas temperature in the plasma chamber, while the rotational temperature characterizes the gas temperature only within the RF power load area.

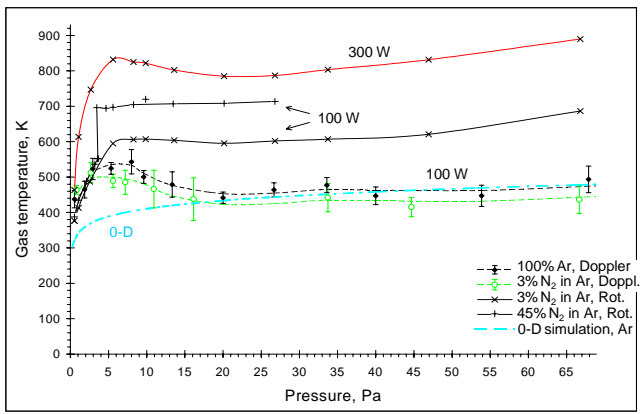
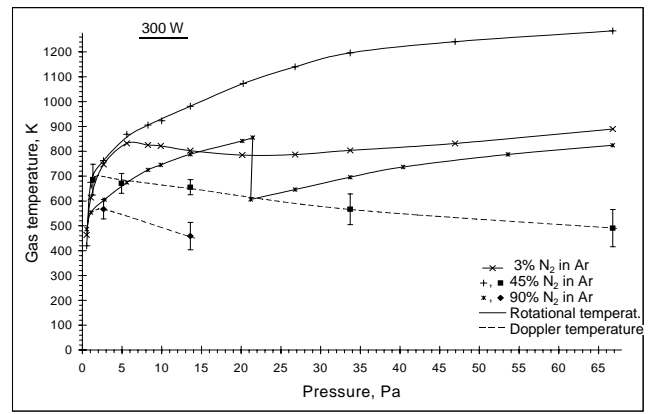

Fig. 5. Gas kinetic temperature (measured and 0-D simulated).   Fig. 6. Gas kinetic temperature measured at 300 W.

Doppler temperatures obtained from 100% argon and 3% $N_2$/Ar mixture were almost identical (Fig. 5) with the temperature in 3% $N_2$/Ar mixture being seemingly slightly lower but the distinction was insignificant at our measurement scatter. Out of the three nitrogen/argon mixtures tested, the 45% $N_2$/Ar mixture exhibited the hottest rotational temperature reaching 1285 K at 300 W at higher pressure (Fig. 6). Note that the same rotational data for 3% $N_2$/Ar mixture at 300 W is plotted in both Figures 5 and 6 to facilitate comparison.

The nitrogen rotational temperature is observed to increase with pressure and power. This rise in temperature with pressure is generally expected in inductively coupled plasmas as gas heating mechanisms involved are proportional to electron-neutral and ion-neutral collision rates, while the dominant cooling mechanisms are only weakly dependent on pressure.[75] However, at large nitrogen concentrations (45, 90%) there are sudden jumps in temperature at 3.5 Pa (100 W, 45% $N_2$) and 21 Pa (300 W, 90% $N_2$) that are associated with a plasma mode transition from inductive to capacitive coupling (see Fig. 5, 6). This plasma mode change is visually apparent as a dimming of the plasma, and is associated with a sharp (order of magnitude) decrease in electron density. The reduction in electron density is accompanied by a reduction in electron-neutral impact frequency and therefore a drop in temperature. However, for the case of 100 W power, the rotational temperature actually increases upon change into capacitive mode, and remains constant as pressure is increased. This likely results from a non-thermal means of populating the emitting $C^3\Pi_u$ state of $N_2$.[76] It is possible for metastable states of argon to transfer energy to ground state nitrogen, thereby populating the $C^3\Pi_u$ state:

$$N_2\left(X^1\Sigma_g^+\right) + Ar\left(3p^5 4s\right) \rightarrow N_2\left(C^3\Pi_u\right) + Ar\left(3p^6\ ^1S_0\right). \tag{2}$$

It has been found that this process results in an elevated rotational temperature in the $C^3\Pi_u$ state of $N_2$. The $C^3\Pi_u$–$B^3\Pi_g$ emission then can be attributed to two different processes, one that does not alter the energy distribution (electron impact) and one that does (excitation transfer from metastable argon). To determine whether or not the measured rotational temperature is valid, we estimated the relative contribution of each excitation mechanism. From the excitation transfer cross sections[76] one can estimate the rate of reaction (2) to be:

$$r_1 = 2.9 \times 10^{-14} \sqrt{\frac{T(K)}{300}}\ n_{Ar^*} n_{N2} \quad cm^3/s. \tag{3}$$

The electron-impact excitation of nitrogen has been parameterized[77] as:

$$r_2 = 1.36 \times 10^{-8} T_e^{0.138} e^{-\frac{11.03}{T_e}} n_e n_{N2} \quad cm^3/s \tag{4}$$

with $T_e$ in eV. For most our conditions, $r_2$ is orders of magnitude larger than $r_1$ and reaction (2) can be considered to be negligible. In the case of Bochkova and Chernysheva,[76] the first reaction was found to be important in the nitrogen afterglow where electron densities and temperatures are low. In our experiments in the capacitive mode at 100 W, this process may be important enough to impact the temperature measurement. More comprehensive measurements, including spatially resolved electron densities and temperatures would be required to ascertain whether or not this process was truly important at our conditions.

The standard deviations of data from Doppler broadening measurements are shown in Figures 5 and 6. The standard deviations of rotational temperature data are omitted for clarity. Typically, random fluctuations of the deduced rotational temperature values were within 1-2% but were the largest (up to ±50 K) at lower pressures for the 3% nitrogen/argon mixture. Overall accuracy of the rotational temperature determination was estimated as limited by ±50 K and was better at higher pressures.

At pressures below 4 Pa, all temperature values measured at 100 W in pure argon and nitrogen/argon mixtures up to 45% $N_2$ agree well within experimental uncertainties (Fig. 5). The same is true for pressures below 2 Pa at 300 W (Fig. 6). This is also in agreement with Hebner,[78] who found no influence of nitrogen addition to argon plasma (up to 50% $N_2$) on gas temperature at 1.3 Pa. The fact that rotational and Doppler temperatures are the same suggests high diffusion rates, and certain similarity in spatial profiles of both thermometric species (Ar* and $N_2$*) at low pressures.

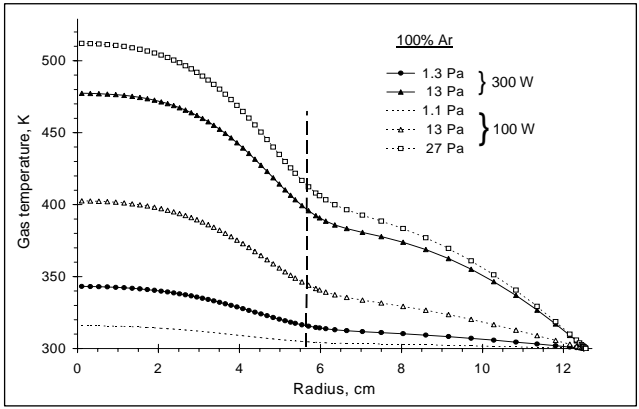
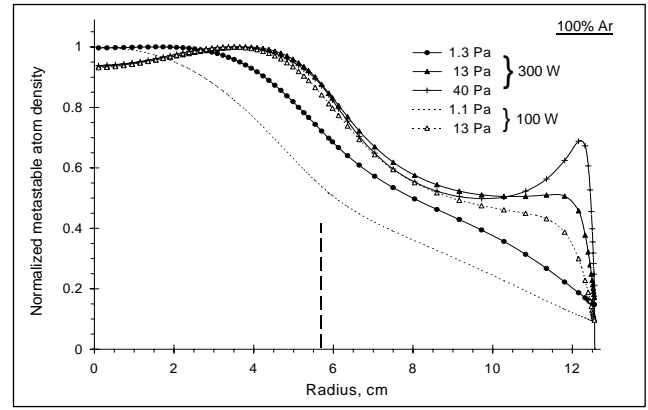

Fig. 7. Simulated radial profiles of gas temperature in Ar plasma.   Fig. 8. Simulated metastable argon profiles in Ar plasma.

As pressure was increased, the rotational and Doppler data diverged from each other indicating more pronounced variations in spatial profiles of the temperature and species. For instance, the measured Doppler temperature was 800 K lower than the rotational temperature at 67 Pa in 45% $N_2$/Ar mixture (Fig. 6). This discrepancy is attributable to the differences in spatial profiles of the thermometric species and therefore, to the differences in effective pathlength, over which temperature was integrated in our experiments.

The origin of discrepancies between measured Doppler and rotational temperatures, attributed to higher spatial gradients in metastable argon atom density at higher pressures, is corroborated by the results of the 0-D simulation of argon plasma (see Fig. 5). The model included direct ionization and excitation by fast electrons, ion-neutral charge exchange, ambipolar diffusion, stepwise ionization, Penning ionization and quenching by electrons and atoms. This model does not predict a local minimum in the pressure dependence of the gas temperature observed around 20 Pa in 100% argon and 3% $N_2$/Ar mixtures for both rotational and Doppler temperatures (Fig. 5). If there were no changes in radial profiles, the gas temperature would be increasing monotonically with pressure as frequency of elastic collisions and ion-to-neutral energy transfer rates increase but diffusion decreases.

Further proof of influence of metastable atom spatial distributions in our plasma reactor on the line-of-sight integrated temperature was obtained with two-dimensional plasma simulation. The radial profiles of gas temperature and metastable argon atoms resulting from the 2-D plasma simulation for a set of pressures and RF powers are plotted versus reactor radius in Figures 7 and 8. Within the range of the simulated conditions (1 - 40 Pa, 100 - 300 W, 100% Ar) the gas temperature profiles were all fairly similar (see Fig. 7). However, the metastable argon atom profiles were changing significantly. At low pressures about 1 Pa, the metastable argon density was maximum in the reactor center. As pressure and power increased, the gas temperature also increased causing depletion of the gas density in the center. Consequently, a higher fraction of metastable argon atoms was pushed out of the center toward the reactor walls (see Fig. 8). Vertical dashed lines in Figures 7 and 8 indicate the position of the outer edge of the induction coil.

Apparently, at low pressures the spatial profiles of both thermometric species (metastable argon atoms and excited nitrogen molecules) have somewhat similar spatial profiles, and thus they yield the same experimental gas temperature values. At higher pressure, the excited nitrogen remains within the glowing area, while metastable argon redistributes significantly and may form local maxima near the reactor wall. Moreover, it might protrude into the window ports of the GEC cell that were not included in the simulated reactor geometry. As a result, the measured Doppler temperature is lower that the rotational temperature at higher pressures.

The effective temperature values (radial-integrated and weighted by the metastable argon density) resulting from 2-D plasma simulation are presented in Fig. 9 in dependence on pressure at 100 and 300 W loaded RF power. Also, the radial-averaged densities of metastable argon atoms are shown. This simulation has captured the correct trend in metastable densities but does not resemble correctly the experimental gas temperature behavior. Probably, the dead space in the reactor window ports that may be filled by metastable argon atoms must be added into the simulation to provide results that would be closer to reality. However, the latter task requires a three-dimensional model. Absolute values of the simulated temperature and densities do not agree well with the experimental data.

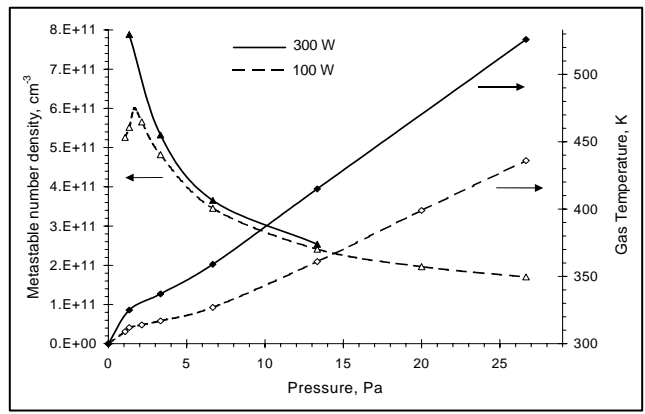

Fig. 9. Simulated gas temperature and metastable atom density in Ar plasma.

Argon absorption recorded throughout our experimental conditions was between 4.5 and 98%. It was the weakest in 90% $N_2$ / 10% Ar mixture and the strongest in 100% argon plasma. Since significantly lower absorption can be reliably measured, small concentrations of argon can always be added to any plasma environment without altering conditions but broadening of their absorption determined. Therefore, excited argon atoms can be used as thermometric species in a wide variety of applications.

Present work indicates the importance of three-dimensional spatial resolution for determining gas temperature gradients in plasma reactors. This can be achieved by measuring laser induced fluorescence. The diode laser induced fluorescence was previously used for analytical purposes.[10-13] Combined with absorption measurements, it can provide spatially resolved absolute densities of species and temperature in the plasma. The plasma chamber should then incorporate three windows (one for the diode laser and two for detection photodiodes). The precise tuning capability of the diode lasers determines very high spectral resolution in fluorescence as well as in absorption. No sophisticated optical equipment is required.

Once understanding of spatial distributions of various species inside the plasma reactor is realized, it is possible to construe a correlation between measured line-of-sight temperature values derived from several different thermometric species and an actual spatial temperature profile. Our work provides an example of determination of the gas temperature only in the central glowing part of the plasma by measuring the nitrogen temperature and an average gas temperature by measuring the argon metastable temperature. The difference in these two values is indicative of the temperature distribution over the reactor volume. An advantage of this approach is simpler hardware requirements in comparison to fluorescence, *i.e.* only two windows necessary.

Our further research will be focused on determination of species in material processing plasmas by continuous-wave cavity ring-down spectroscopy. Only a few recent research papers have dealt with CRDS in various plasmas.[79-82] None of them were applied in a cw mode (generally more sensitive with higher spectral resolution as compared to the conventional pulsed CRDS), nor used the diode lasers that are compact and inexpensive. The detection limit for $C_nH_m$ radicals is estimated at $\sim 10^9$ cm$^{-3}$; and at $\sim 10^{11}$ cm$^{-3}$ for $CF_X$ radicals.[40] This level of sensitivity is sufficient for diagnostic applications in plasma-based processing.

## 5. CONCLUSIONS

Application of a miniature diode laser based sensor for determination of the gas temperature inside an inductively coupled plasma reactor was demonstrated. The results were compared to rotational temperature determination at the same conditions and to the gas temperature calculated using a 2-D model. These comparisons revealed strongly non-uniform and variable thermal spatial distributions depending on conditions.

The thermometric species used in this study were metastable argon atoms and nitrogen molecules. In principle, any species with absorption features within the tunability range of the available laser can serve as thermometric. Alternatively, a small concentration of argon added to any other plasma will not alter the process but enable the gas temperature determination by argon absorption.

The multiplexing diode laser based absorption, enhanced in a modulated high-reflectivity optical resonator is suggested as an ideal and almost universal technique for acquisition of the real-time *in situ* multi-parametric data (species densities and temperature), which makes it possible to determine absolute concentrations of atoms or radicals and their temperature, thus providing a single, simple, compact, and inexpensive tool to optimize and control reproducible micro/nano-fabrication processes with active sensor-driven feedback. A laser beam can be adjusted to analyze the local reactions occurring in the vicinity of the designed nano-scale features.

In principle, sharp three-dimensional spatial resolution can be achieved with this technology if laser induced fluorescence is recorded simultaneously. Excellent spectral resolution is determined by the spectral width of laser radiation (below $1 \times 10^{-4}$ nm).

Precision of temperature measurements can be further perfected to a few Kelvin. The lower limits for determination of $C_nH_m$ radicals by enhanced absorption in a modulated optical cavity are estimated at $\sim 10^9$ cm$^{-3}$; and for $CF_X$ radicals at $\sim 10^{11}$ cm$^{-3}$.

## ACKNOWLEDGMENTS

The presented results were obtained when A.A.B. held a National Research Council associateship award at NASA Ames Research Center. The work of B.A.C. was funded by NASA Ames contract NAS2-99092 to Eloret Corporation. The authors are thankful to David B. Hash and Gabriel J. Laden for their assistance in 2-D plasma simulation.

# REFERENCES

1. G.F. Cerofolini, G. Ferla, "Toward a hybrid micro-nanoelectronics." *J. Nanoparticle Res.*, **4**, 185-191 (2002).
2. *International Technology Roadmap for Semiconductors, 2003 Eddition*. Available: *http://public.itrs.net*
3. R. Compañó, Ed., *Technology Roadmap for Nanoelectronics, 2nd ed.*, European Commission, IST programme, 2000. Available: *www.cordis.lu/ist/fet/nidqf.htm*
4. K.L. Wang, "Issues of nanoelectronics: a possible roadmap." *J. Nanosci. Nanotech.*, **2**, 235-266 (2002).
5. J.A. Hutchby, G.I. Bourianoff, V.V. Zhirnov, J.E. Brewer, "Extending the road beyond CMOS." *IEEE Circ. Devices Mag.*, **18**, No.2, 28-41 (2002).
6. B.N. Khare, M. Meyyappan, A.M. Cassell, C.V. Nguyen, J. Han, "Functionalization of carbon nanotubes using atomic hydrogen from a glow discharge." *Nano Lett.*, **2**, 73-77 (2002).
7. Q. Chen, L. Dai, M. Gao, S. Huang, A. Mau, "Plasma activation of carbon nanotubes for chemical modification." *J. Phys. Chem. B*, **105**, 618-622 (2001).
8. D.S. Baer, R.K. Hanson, "Tunable diode laser absorption diagnostics for atmospheric pressure plasmas." *J. Quant. Spectrosc. Radiat. Transfer*, **47**, 455-475 (1992).
9. B.A. Paldus, R.N. Zare, "Absorption spectroscopies: from early beginnings to cavity-ringdown spectroscopy." Chap. 5 in *Cavity-Ringdown Spectroscopy*, Eds. K.W. Busch, M.A. Busch, Am. Chem. Soc., Washington, 1999, 49-70.
10. P.E. Walters, T.E. Barber, M.W. Wensing, J.D. Winefordner, "A diode laser wavelength reference system applied to the determination of rubidium in atomic fluorescence spectroscopy." *Spectrochim. Acta, Part B*, **46**, 1015-1020 (1991).
11. A. Zybin, C. Schnürer-Patschan, K. Niemax, "Simultaneous multielement determination in a commercial graphite furnace by diode laser induced fluorescence." *Spectrochim. Acta, Part B*, **47**, 1519-1524 (1992).
12. K. Yuasa, K. Yamashina, T. Sakurai, "Ar lowest excited state densities in Ar and Ar-Hg hot cathode discharge." *Jpn. J. Appl. Phys.*, **36**, 2340-2345 (1997).
13. C. Raab, J. Bolle, H. Oberst, J. Eschner, F. Schmidt-Kaler, R. Blatt, "Diode laser spectrometer at 493 nm for single trapped $Ba^+$ ions." *Appl. Phys. B*, **67**, 683-688 (1998).
14. W. Wang, M.M. Fejer, R.H. Hammond, M.R. Beasley, C.H. Ahn, "Atomic absorption monitor for deposition process control of aluminum at 394 nm using frequency-doubled diode laser." *Appl. Phys. Lett.*, **68**, 729-731 (1996).
15. W. Wang, R.H. Hammond, M.M. Fejer, M.R. Beasley, "Atomic flux measurement by diode-laser-based atomic absorption spectroscopy." *J. Vac. Sci. Technol. A*, **17**, 2676-2684 (1999).
16. P. Kluczynski, Å.M. Lindberg, O. Axner, "Background signals in wavelength-modulation spectroscopy by use of frequency-doubled diode-laser light. II. Experiment." *Appl. Opt.*, **40**,794-805 (2001).
17. A. Zybin, G. Schaldach, H. Berndt, K. Niemax, *Anal. Chem.*, **70**, 5093-5096 (1998).
18. K.B. MacAdam, A. Steimbach, C. Wieman, "A narrow-band tunable diode laser system with grating feedback and a saturated absorption spectrometer for Cs and Rb." *Am. J. Phys.*, **60**, 1098-1111 (1992).
19. T. Laurila, R. Hernberg, "Frequency-doubled diode laser for ultraviolet spectroscopy at 325 nm." *Appl.Phys.Lett.*, **83**, 845-847 (2003).
20. E.C. Jung, K.-H. Ko, S.P. Rho, C. Lim, C.-J. Kim, "Measurement of the populations of metastable levels in gadolinium vapor by diode laser-based UV and near-IR absorption spectroscopy." *Opt. Commun.*, **212**, 293-300 (2002).
21. J. Alnis, U. Gustafsson, G. Somesfalean, S. Svanberg, *Appl. Phys. Lett.*, **76**, 1234-1236 (2000).
22. R.F. Tale, B.T. Anderson, P.B. Keating, G.D. Hager, "Diode-laser Zeeman spectroscopy of atomic iodine." *J. Opt. Soc. Am. B*, **17**, 1271-1278 (2000).
23. H. Leinen, D. Gläßner, H. Metcalf, R. Wynands, D. Haubrich, D. Meschede, "GaN blue diode lasers: a spectroscopist's view." *Appl. Phys. B*, **70**, 567-571 (2000).
24. U. Gustafsson, G. Somesfalean, J. Alnis, S. Svanberg, "Frequency-modulation spectroscopy with blue diode lasers." *Appl. Opt.*, **39**, 3774-3780 (2000).
25. C. Schnürer-Patschan, A. Zybin, H. Groll, K. Niemax, "Improvement in detection limits in graphite-furnace diode-laser atomic-absorption spectrometry by wavelength modulation technique - plenary lecture." *J. Anal. At. Spectrom.*, **8**,1103-1107 (1993).
26. H.D. Wizermann, K. Niemax, *Spectrochim. Acta, Part B*, **55**, 637-650 (2000).
27. D.J. Butcher, A. Zybin, M.A. Bolshov, K. Niemax, *Anal. Chem.*, **71**, 5379-5385 (1999).
28. H. Park, M. Lee, E.C. Jung, J. Yi, Y. Rhee, J. Lee, "Isotope shifts of Sm I measured by diode-laser-based Doppler-free spectrometry." *J. Opt. Soc. Am. B*, **16**, 1169-1174 (1999).
29. R.J. Lipert, S.C. Lee, M.C. Edelson, "Application of diode lasers to actinide atom monitoring." *Appl. Spectrosc.*, **46**, 1307-1308 (1992).

30. W. Wang, R.H. Hammond, M.M. Fejer, C.H. Ahn, M.R. Beasley, "Diode-laser-based atomic absorption monitor using frequency-modulation spectroscopy for physical vapor deposition process control." *Appl. Phys. Lett.*, **67**, 1375-1377 (1995).
31. E. Augustyniak, S. Filimonov, C.-S. Lu, "Investigation of thermalization process in sputtering systems by atomic absorption spectroscopy." *Proc. SPIE*, **3507**, 192-200 (1998).
32. D.J. Butcher, A. Zybin, M.A. Bolshov, K. Niemax, "Diode laser atomic absorption spectroscopy as a detector for metal speciation." *Rev. Anal. Chem.*, **20**, 79-100 (2001).
33. D.C. Hovde, J.T. Hodges, G.E. Scace, J.A. Silver, "Wavelength-modulation laser hygrometer for ultrasensitive detection of water vapor in semiconductor gases." *Appl. Opt.*, **40**, 829-839 (2001).
34. B.L. Upschulte, D.M. Sonnenfroh, M.G. Allen, "Measurements of CO, $CO_2$, OH, and $H_2O$ in room-temperature and combustion gases by use of a broadly current-tuned multisection InGaAsP diode laser." *Appl. Opt.*, **38**, 1506-1512 (1999).
35. J. Wang, M. Maiorov, D.S. Baer, D.Z. Garbuzov, J.C. Connolly, R.K. Hanson, "*In situ* combustion measurements of CO with diode-laser absorption near 2.3 μm." *Appl. Opt.*, **39**, 5579-5589 (2000).
36. M.E. Webber, S. Kim, S.T. Sanders, D.S. Baer, R.K. Hanson, Y. Ikeda, "*In situ* combustion measurements of $CO_2$ by use of a distributed-feedback diode-laser sensor near 2.0 μm." *Appl. Opt.*, **40**, 821-828 (2001).
37. D.C. Hovde, A.C. Stanton, T.P. Meyers, D.R. Matt, "Methane emissions from a landfill measured by eddy correlation using a fast response diode laser sensor." *J. Atmos. Chem.*, **20**, 141-162 (1995).
38. A. Ubukata, J. Dong, H. Masusaki, T. Saton, K. Matsumoto, "Hydrogen chloride gas monitoring at 1.74 μm with InGaAs/InGaAsP strained quantum well laser." *Jpn. J. Appl. Phys.*, **37**, 2521-2523 (1998).
39. S.-I Chou, D.S. Baer, R.K. Hanson, W.Z. Collison, T.Q. Ni, "HBr concentration and temperature measurements in a plasma etch reactor using diode laser absorption spectroscopy." *J. Vac. Sci. Technol. A*, **19**, 477-484 (2001).
40. A.A. Bol'shakov, S.P. Sharma, M. Meyyappan, "Detection of fluorocarbons $CF_X$ in a semiconductor etching ICP reactor by absorption spectroscopy." *ICP Inf. Newslett.*, **27**, no.11, 784-785 (2002).
41. L. Delzeit, I. McAninch, B.A. Cruden, D. Hash, B. Chen, J. Han, M. Meyyappan, "Growth of multiwall carbon nanotubes in an inductively coupled plasma reactor." *J. Appl. Phys.*, **91**, 6027-6033 (2002).
42. M.E. Littau, M.J. Sowa, J.L. Cecchi, "Diode laser measurements of $CF_X$ species in a low-pressure, high-density plasma reactor." *J. Vac. Sci. Technol. A*, **20**, 1603-1610 (2002).
43. J. Wormhoudt, Radical and moleculer product concentration measurements in $CF_4$ and $CH_4$ radio-frequency plasmas by infrared tunable diode-laser absorption." *J. Vac. Sci. Technol. A*, **8**, 1722-1725 (1990).
44. K. Miyata, M. Hori, T. Goto, "$CF_X$ (x = 1–3) radical densities during Si, $SiO_2$, and $Si_3N_4$ etching employing electron cyclotron resonance $CHF_3$ plasma." *J. Vac. Sci. Technol. A*, **15**, 568-572 (1997).
45. P.B. Davies, D.M. Smith, "Diode-laser spectroscopy and coupled analysis of the $\nu_2$ and $\nu_4$ fundamental bands of $SiH_3^+$." *J. Chem Phys.*, **100**, 6166-6174 (1994).
46. H.C. Sun, E.A. Whittaker, "Real-time in-situ detection of $SF_6$ in a plasma reactor." *Appl. Phys. Lett.*, **63**, 1035-1037 (1993).
47. S. Kim, D.P. Billesbach, R. Dillon, "Tunable diode laser spectroscopy measurement of $CH_3$ and $C_2H_2$ densities in a $H_2O/CH_3OH$ radio frequency chemical vapor deposition diamond system." *J. Vac. Sci. Technol. A*, **15**, 2247-2251 (1997).
48. A.I. Nadezhdinskii, E.V. Stepanov, I.I. Zasavitskii, "Spectral gas analysis of polyatomic molecules by tunable diode lasers." *Proc. SPIE*, **1724**, 238-250 (1992).
49. J. Wormhoudt, A.C. Stanton, A.D. Richards, H.H. Sawin, "Atomic chlorine concentration and gas temperature measurements in a plasma-etching reactor." *J. Appl. Phys.*, 61, 142-149 (1987).
50. I.Yu. Baranov, V.I. Demidov, N.B. Kolokolov, "Temperature dependence of rate constants for metastable atomic-argon deactivation by slow electrons." *Opt. Spektrosk.*, **51**, 571-574 (1981) [Engl. transl.: *Opt. Spectrosc.*, **51**, 316-318 (1981)].
51. C.M. Ferreira, J. Loureiro, A. Ricard, "Populations in the metastable and the resonance levels of argon and stepwise ionization effects in a low-pressure argon positive column." *J. Appl. Phys.*, **57**, 82-90 (1985).
52. D. Leonhardt, C.R. Eddy, V.A. Shamamian, R.F. Fernsler, J.E. Butler, "Argon metastables in a high density processing plasma." *J. Appl. Phys.*, **83**, 2971-2979 (1998).
53. M.I. Boulos, P. Fauchais, E. Pfender, *Thermal Plasmas. Fundamentals and Applications*, Plenum Press, New York, 1994.
54. K. Tachibana, H. Harima, Y. Urano, "Measurements of collisional broadening and the shift of argon spectral lines using a tunable diode laser." *J. Phys. B*, **15**, 3169-3178 (1982).
55. D.S. Baer, R.K. Hanson, "Tunable diode laser absorption diagnostics for atmospheric pressure plasmas." *J. Quant. Spectrosc. Radiat. Transfer*, **47**, 455-475 (1992).
56. J.M. de Regt, R.D. Tas, J.A.M. van der Mullen, "A diode laser absorption study on a 100 MHz argon ICP." *J. Phys. D*, **29**, 2404-2412 (1996).

57. N. Beverini, G.D. Gobbo, G.L. Genovesi, F. Maccarrone, F. Strumia, F. Paganucci, A. Turco, M. Andrenucci, "Time-resolved plasma diagnostic by laser-diode spectroscopy." *IEEE J. Quant. Electron.*, **32**, 1874-1881 (1996).
58. I.A. Porokhova, Yu.B. Golubovskii, C. Csambal, V. Helbig, C. Wilke, J.F. Behnke, "Nonlocal electron kinetics and excited state densities in a magnetron discharge in argon." *Phys. Rev. E*, **65**, 046401/1-10 (2002).
59. M. Suzuki, K. Katoh, N. Nishimiya, "Saturated absorption spectroscopy of Xe using a GaAs semiconductor laser." *Spectrochim. Acta, Part A*, **58**, 2519-2531 (2002).
60. R. Uhl, J. Franzke, U. Haas, "Detection of argon and krypton traces in noble gases by diode laser absorption spectrometry." *Appl. Phys. B*, **73**, 71-74 (2001).
61. N. Sadeghi, M. van de Grift, D. Vender, G.M.W. Kroesen, F.J. de Hoog, "Transport of argon ions in an inductively coupled high-density plasma." *Appl. Phys. Lett.*, **70**, 835-837 (1997).
62. M. Miclea, K. Kunze, G. Musa, J. Franzke, K. Niemax, *Spectrocim. Acta, Part B*, **56**, 37-43 (2001).
63. D.S. Baer, H.A. Chang, R.K. Hanson, "Semiconductor laser absorption diagnostics of atomic oxygen in an atmospheric-pressure plasma." *J. Quant. Spectrosc. Radiat. Transfer*, **50**, 621-633 (1993).
64. H.P. Zappe, M. Hess, M. Moser, R. Hövel, K. Gulden, H.-P. Gauggel, F.M. di Sopra, "Narrow-linewidth vertical-cavity surface-emitting lasers for oxygen detection." *Appl. Opt.*, **39**, 2475-2479 (2000).
65. S.T. Sanders, J. Wang, J.B. Jeffries, R.K. Hanson, "Diode-laser absorption sensor for line-of-sight gas temperature distributions." *Appl. Opt.*, **40**, 4404-4415 (2001).
66. J. Wang, S.T. Sanders, J.B. Jeffries, R.K. Hanson, "Oxygen measurements at high pressures with vertical cavity surface-emitting lasers." *Appl. Phys. B*, **72**, 865-872 (2001).
67. E. Schlosser, T. Fernholz, H. Teichert, V. Ebert, "In situ detection of potassium atoms in high-temperature coal-combustion systems using near-infrared-diode lasers." *Spectrochim. Acta, Part A*, **58**, 2347-2359 (2002).
68. J.A. Silver, D.J. Kane, "Diode laser measurements of concentration and temperature in microgravity combustion." *Meas. Sci. Technol.*, **10**, 845-852 (1999).
69. J.S. Kim, M.V.V.S. Rao, M.A. Cappelli, S.P. Sharma, M. Meyyappan, "Mass spectrometric and Langmuir probe measurements in inductively coupled plasmas in Ar, $CHF_3/Ar$ and $CHF_3/Ar/O_2$ mixtures." *Plasma Sources Sci. Technol.*, **10**, 191-204 (2001).
70. B.A. Cruden, M.V.V.S. Rao, S.P. Sharma, M. Meyyappan, "Neutral gas temperature estimates in an inductively coupled $CF_4$ plasma by fitting diatomic emission spectra." *J. Appl. Phys.*, **91**, 8955-8964 (2002).
71. F. Roux, F. Michaud, M. Vervloet, "High-resolution Fourier spectrometry of $^{14}N_2$ violet emission spectrum: extensive analysis of the $C^3\Pi_u$ from $B^3\Pi_g$ system." *J. Mol. Spec.*, **158**, 270-277 (1993).
72. J.M. Brown, A.J. Merer, "Lambda-type doubling parameters for molecules in Pi electronic states in triplet and higher multiplicity." *J. Mol. Spec.*, **74**, 488-494 (1979).
73. S.-L. Cheah, Y.-P. Lee, J.F. Ogilvie, "Wavenumbers, strengths, widths and shifts with pressure of lines in four bands of gaseous $^{16}O_2$ in the systems $a^1\Delta_g - X^3\Sigma_g^-$ and $b^1\Sigma_g - X^3\Sigma_g^-$." *J. Quant. Spectrosc. Radiat. Transfer*, 64, 467-482 (2000).
74. S. Nowak, J.A.M. van der Mullen, B. van der Sijde, D.C. Schram, "Spectroscopic determination of electron density and temperature profiles in an inductively-coupled argon plasma." *J. Quant. Spectrosc. Radiat. Transfer*, **41**, 177-186 (1989).
75. B.A. Cruden, M.V.V.S. Rao, S.P. Sharma, M. Meyyappan, "Neutral gas temperature estimate in $CF_4/O_2/Ar$ inductively coupled plasmas." *Appl. Phys. Lett.*, **81**, 990-992 (2002).
76. O.P. Bochkova, N.V. Chernysheva, "Excitation of nitrogen molecules in a high-frequency argon-nitrogen discharge." *Opt. Spektrosk.*, **31,** 677-681 (1971). [Engl. transl.: *Opt. Spectrosc.*, **31**, 359-361 (1971)].
77. M. Capitelli, C.M. Ferreira, B.F. Gordiets, A.I. Osipov, *Plasma Kinetics in Atmospheric Gases*, Springer-Verlag, New York, 2000.
78. G.A. Hebner, "Spatially resolved, excited state densities and neutral and ion temperatures in inductively coupled argon plasmas." *J. Appl. Phys.*, **80**, 2624-2636 (1996).
79. J.P. Booth, G. Cunge, L. Biennier, D. Romanini, A. Kachanov, "Ultraviolet cavity ring-down spectroscopy of free radicals in etching plasmas." *Chem. Phys. Lett.*, 317, 631-636 (2000).
80. F. Grangeon, C. Monard, J.-L. Dorier, A.A. Howling, C. Hollenstein, D. Romanini, N. Sadeghi, "Applications of the cavity ring-down technique to a large-area rf-plasma reactor." *Plasma Sources Sci. Technol.*, **8**, 448-456 (1999).
81. E. Quandt, I. Kraemer, H.F. Döbele, "Measurements of negative-ion densities by cavity ringdown spectroscopy." *Europhys. Lett.*, **45**, 32-37 (1998).
82. C. Wang, F. Mazzotti, G. Miller, C. Winstead, "Cavity ringdown spectroscopy for diagnostic and analytical measurements in an inductively coupled plasma." *Appl. Spectrosc.*, **56**, 386-397 (2002).